\documentclass[superscriptaddress,pre,twocolumn]{revtex4-1}
\usepackage{graphicx}
\usepackage{amsmath}
\usepackage{amsfonts}
\usepackage{amssymb}
\usepackage{multirow}
\usepackage{colortbl}
\usepackage{xcolor}
\usepackage{hhline}
\usepackage{longtable}
\newcommand{\be}{\begin{equation}}
\newcommand{\ee}{\end{equation}}
\newcommand{\ep}{\epsilon}
\newcommand{\R}{\mathcal{R}}

\newcommand{\gr}{\cellcolor{lightgray}}
\newcommand{\T}{\mathcal{A}^{(0)}}

\begin{document}

\title{Semiclassical calculation of spectral correlation functions of chaotic systems}
\author{Sebastian M\"uller}
\affiliation{School of Mathematics, University of Bristol, Bristol BS8 1TW, UK}
\author{Marcel Novaes}
\affiliation{Instituto de F\'isica, Universidade Federal de Uberl\^andia, Uberl\^andia,
MG, 38408-100, Brazil}

\begin{abstract}

We present a semiclassical approach to $n$-point spectral correlation functions of quantum systems
whose classical dynamics is chaotic, for arbitrary $n$. The basic ingredients are sets of periodic orbits
that have nearly the same action and therefore provide constructive interference. We
calculate explicitly the first correlation functions, to leading orders in their energy
arguments, for both unitary and orthogonal symmetry classes. The results agree with corresponding 
predictions from random matrix theory, thereby giving solid support to the conjecture of universality.

\end{abstract}

\maketitle

\section{Introduction}

According to the Bohigas-Giannoni-Schmit (BGS) conjecture, put forward 30 years ago \cite{bgs}
(see also \cite{bgs2,bgs3}), highly excited energy levels of generic chaotic systems have
universal local spectral statistics. This universality is captured by the random matrix
theory (RMT) approach to quantum chaos: such statistics are expected to agree with those
of the Gaussian Ensembles, the particular ensemble (unitary, orthogonal or symplectic)
being determined by the overall symmetries of the system \cite{haake}. For spinless particles, 
the unitary class corresponds to systems with broken time-reversal symmetry (TRS), while the 
orthogonal class corresponds to systems with preserved TRS.

Obtaining spectral statistics for a specific system should be possible within the
semiclassical approximation, by using the periodic orbit theory of the Gutzwiller trace
formula \cite{gutz} and ergodic properties of long orbits. Indeed, progress in this
direction was made early on in \cite{hoda,berry}, deriving the leading term in
$(\ep_1-\ep_2)^{-1}$ of the $2$-point spectral correlation function $\R_2(\ep_1-\ep_2)$ by
considering only interference of an orbit with itself (so-called `diagonal
approximation'). For time-reversal invariant systems also interference between
mutually time-reversed orbits was taken into account.

Interference between orbits which are not identical up to time reversal
was expected to give higher-order contributions, as conjectured and supported
numerically in \cite{argaman}. This interference  started to be accounted for
perturbatively in the work of Sieber and Richter \cite{sieber,sieber2}, providing the next-to-leading term of $\R_2$. It
was suggested that the mechanism producing systematic interference is the existence of
`encounters' in long orbits, regions of phase space where the orbit comes very close to
itself, up to time reversal. This theory allowed the calculation of $\R_2$
for all universality classes, at all orders of perturbation theory \cite{R2a,R2b,R2c} and
even beyond perturbation theory \cite{longa,km,longb}.

However, it is the semiclassical derivation of all $n$-point correlation functions,
denoted $\R_n$, which would embody the full BGS conjecture. This has so far remained a
challenge  
because it involves multiplets of correlated periodic orbits with $n$ different energies.
An exception is \cite{taro} where 
a variant of the diagonal approximation also accounting for non-perturbative effects was evaluated for the unitary symmetry class, however leaving out perturbative contributions due to encounters.
(See also \cite{shukla} for the standard diagonal approximation.)
Interestingly, the analogous problem has evolved more rapidly in the transport setting
\cite{combinat1,combinat2,combinat3,combinat4,combinat6,combinat7,combinat5,combinat8},
where the semiclassical calculation of counting statistics requires multiplets of
correlated scattering trajectories, but all at the same energy (but energy correlations in scattering have also been considered \cite{energy1,energy2}). We also note that for quantum graphs an understanding of higher-order correlation functions has been achieved in \cite{weidenmueller1, weidenmueller2}. 

In this work we present substantial progress in the semiclassical calculation of
$\R_n$, both for systems with and without TRS. First, we derive a set of diagrammatic rules that reduces the problem to the
counting of certain diagrams which in turn, following previous works \cite{R2c,combinat6}, we relate to factorizations of permutations. This allows, in principle, any finite order in perturbation
theory to be obtained for all symmetry classes. We then present the explicit calculation
of the leading orders for the first correlation functions, and show that the results
agree with corresponding predictions from random matrix theory.

\section{Spectral correlation functions}

\subsection{Definition}

Let $\rho(E)$ denote the system's density of states. This can be divided into the smooth
Weyl part $\bar{\rho}$, related to the volume of the energy shell in phase space, and a fluctuating
part, $ \rho(E)= \bar\rho(E)+\rho_{\rm fl}(E).$ Then \be\label{Rndef} \R_n(\ep)=
\frac{1}{\bar\rho^n}\left\langle
\prod_{i=1}^n\rho\left(E+\frac{\ep_i}{\bar\rho}\right)\right\rangle \ee is the $n$-point
correlation function expressed in terms of dimensionless energy differences $\epsilon_i$. The brackets denote an average with respect to $E$, over an interval small enough to neglect
variations of $\bar\rho$.

It is easy to relate $\R_n(\epsilon)$ to a similar quantity, defined in terms of only the
fluctuating part of the spectral density, \be\label{Rn} \widetilde\R_n(\epsilon)=
\frac{1}{\bar\rho^n}\left\langle \prod_{i=1}^n\rho_{\rm
fl}\left(E+\frac{\epsilon_i}{\bar\rho}\right)\right\rangle.\ee Since by definition $\rho_{\rm fl}$ has a
vanishing average, we obtain $\widetilde\R_{1}=0$. Thus, we have, for example,
\be\label{R2til} \R_2(\ep_1,\ep_2)=1+\widetilde\R_2(\ep_1,\ep_2)\ee and \be\label{R3til}
\R_3(\ep_1,\ep_2,\ep_3)=1+\sum_{j=1}^3\sum_{k>j}^3\widetilde\R_2(\ep_j,\ep_k)+\widetilde\R_3(\ep_1,\ep_2,\ep_3).
\ee

\subsection{Results from random matrix theory}

Correlation functions for the Gaussian ensembles have long been studied within random matrix theory, using different
methods (see, e.g. \cite{Mehta,Forrester,brezin,kanzieper}). In the regime of large matrices, the calculation of $\R_n$ for the Gaussian
Unitary Ensemble reduces to the calculation of $n\times n$ determinants, \be
\R_n(\ep)=\det\left(\frac{\sin[\pi(\ep_i-\ep_j)]}{\pi(\ep_i-\ep_j)}\right).\ee For
example, \begin{align}
\R_2(\ep_1,\ep_2)&=1-\left(\frac{\sin[\pi(\ep_1-\ep_2)]}{\pi(\ep_1-\ep_2)}\right)^2
\\&=1+\frac{2}{\pi^2(\ep_1-\ep_2)^2}-\frac{2\cos[2\pi(\ep_1-\ep_2)]}{\pi^2(\ep_1-\ep_2)^2}.\end{align}
In general, $\R_n(\ep)$ can be divided into \emph{oscillatory} and \emph{non-oscillatory}
terms, the former containing trigonometric functions of the variables $\ep_i-\ep_j$, and
the latter being a Laurent polynomial in these variables.

The semiclassical approach we employ here, based on the Gutzwiller trace formula, is only
able to address the non-oscillatory terms. More refined approaches, based on the
so-called Riemann-Siegel look-alike formula, have been employed in order to derive the
oscillatory terms. We believe the approach presented here may be adapted to this more general
setting.

Systems with time-reversal symmetry are modeled by the Gaussian Orthogonal Ensemble. In that case the correlation functions are expressible as $2n\times 2n$ Pfaffians, \be\label{Pf} \R_n(\ep)={\rm Pf}\left(\begin{matrix}
D(\ep_i-\ep_j)&S(\ep_i-\ep_j)\\-S(\ep_i-\ep_j)&I(\ep_i-\ep_j)\end{matrix}\right),\ee
Here the matrix consists of $2\times2$ blocks labelled by $i,j=1\ldots n$, and the entries involve the functions   \be S(x)=\frac{\sin(\pi x)}{\pi x}\ee and \be
D(x)=\int_0^1du\, u \sin(\pi u x),\quad I(x)=-\int_1^\infty \frac{du}{u} \sin(\pi u x).\ee

These correlations can also be divided into oscillatory and non-oscillatory terms, but
the latter are now infinite series in the variables $(\ep_i-\ep_j)^{-1}$. For example,
the non-oscillatory terms for $\R_2$ in this class are \be \R_2^{\rm no}(\ep_1,\ep_2)=
1-\frac{1}{\pi^{2}(\ep_1-\ep_2)^2}+\frac{3}{2{\pi }^{4}(\ep_1-\ep_2)^4}+\cdots\ee
 Explicit formulas for $n>2$ will be given later, see Eqs. (\ref{goe3}) and (\ref{R4}).

\subsection{Semiclassical Approximation}

A semiclassical approach to the problem is justified since we consider high-lying states.
This must start from the celebrated Gutzwiller trace formula \cite{haake,gutz}, which
asymptotically as $\hbar\to 0$ relates $\rho_{\rm fl}$ to the isolated and unstable
periodic orbits of the classical dynamics: \be\label{gutz} \rho_{\rm
fl}(E)\approx\frac{1}{\pi\hbar}{\rm Re}\sum_pF_pT_p^{\rm prim}e^{iS_p(E)/\hbar},\ee where $S_p$ and
$T_p^{\rm prim}$ are the action and primitive period of the orbit $p$, respectively, while $F_p$ is a
stability factor. (Following the notaton of \cite{longb}, this involves   the monodromy matrix and includes the complex Maslov phase
factor). The primitive period coincides with the period $T_p$, unless the
orbit involves multiple repetitions of a shorter orbit, in which case the
period of the shorter orbit has to be used. However, for any given range of
periods
the number of  orbits involving repetitions of shorter ones is negligible compared to
the overall number of orbits, hence we can  replace $T_p^{\rm prim}$  by $T_p$ without affecting the final result of our theory.
Making this replacement and introducing the Heisenberg time 
\be
T_H=2\pi\hbar\bar\rho\ee 
we thus write (\ref{gutz}) as
 \be\label{gutz2} \rho_{\rm
fl}(E)\approx\frac{1}{T_H}\Big(\sum_pF_pT_p e^{iS_p(E)/\hbar}+{\rm c.c.}\Big).\ee

To compute correlation functions, the infinite sum (\ref{gutz2}) must be inserted into
$\widetilde\R_n$, leading to multiple sums over periodic orbits.   For example,
$\widetilde\R_2(\ep_1,\ep_2)$ is approximated by \be
\frac{2}{T_{H}^2}{\rm Re}\left\langle
\sum_{p,q}F_pF^*_qT_pT_qe^{i(S_p(E_1)-S_q(E_2))/\hbar}\right\rangle,\ee where
$E_j=E+\ep_j/\bar\rho$.
Crucially, the contributions from most pairs of orbits will oscillate rapidly
as the energy is varied and are washed out by the energy average. 
This can only be avoided if the action difference is small, giving the exponent  a chance of being
stationary with respect to small changes in $E$.

In general, inserting the trace formula (\ref{gutz2}) into $\widetilde\R_n(\ep)$,
Eq. (\ref{Rn}),
we obtain a multiple sum over periodic orbits. For each orbit sum we have to take into account the possibility that the action
appears with a positive sign (corresponding to the initial term in Eq.(\ref{gutz2})), or  with a negative
sign (corresponding to the complex conjugated term).
It is thus useful to split $\widetilde\R_n(\ep)$ into contributions $\widetilde\R_{J,K}(\epsilon)$
where $J$ actions contribute with a positive sign, and $K=n-J$ actions contribute with a negative sign.
In terms of these auxiliary quantities, we have \be\label{Rsum} \widetilde\R_n(\ep)=\sum_{J=0}^n
\widetilde\R_{J,n-J}(\ep)\ee

To obtain a concrete formula for $\widetilde\R_{J,K}(\epsilon)$ we first consider the case that the first $J$ actions, associated to the energy increments $\epsilon_1,\ldots,\epsilon_J$ contribute with a positive sign. We denote the corresponding orbits by $p_1,\ldots,p_J$ and assemble them into a set $P$.
The remaining $K$ actions, associated to increments $\epsilon_{J+1},\ldots,\epsilon_n$ and orbits from $Q=\{q_1,\ldots,q_K\}$, contribute with a negative sign.
With the notation
\be
\label{epsilon_eta}
\eta_k=\epsilon_{J+k}.
\ee
for the latter increments we obtain
\be\label{RJK}
R_{J,K}(\ep,\eta)=\frac{1}{T_H^n}\left\langle\sum_{P,Q}F_PF^*_QT_PT_Qe^{i\Delta
S/\hbar}\right\rangle,\ee with the action difference \be \Delta
S=\sum_{j=1}^JS_{p_j}(E+\epsilon_j/\bar\rho)-\sum_{k=1}^KS_{q_k}(E+\eta_k/\bar\rho).\ee In order to write the multiple sums in compact form, we have also  defined collective
stability
factors and period products as $F_P=\prod_{p\in P}F_p$ and $T_P=\prod_{p\in P}T_p$.

To obtain $\widetilde\R_{J,K}(\epsilon)$ we have to consider all ways of
splitting the $n$ orbits into $J$ orbits for which the action is taken with a positive sign and $K$ orbits where it is taken with a negative sign. This can be done by summing over all permutations of the $n$ energy increments. This operation turns $\widetilde\R_n$ into a symmetric function of all energy increments;
in the following it
will be denoted by the operator ${\rm Sym}_n$.
However summing over all permutations is actually too much as this also involves exchanges among the $J$ energy increments associated to positive signs in the exponent, and among the $K$ increments associated to negative signs. To compensate this we have to divide out the number $J!K!$ if such exchanges and write 
 \be\label{sym} \widetilde\R_{J,K}(\ep)=\frac{1}{J!K!}{\rm Sym}_n
\Big[R_{J,K}(\ep,\eta)\Big],\ee 

Given $\widetilde\R_{J,K}(\ep)$ we can then access $\widetilde\R_n(\epsilon)$
using (\ref{Rsum}). Here, the extreme cases $J=0$ and $J=n$ are not
allowed in practice, as in these cases all orbits contribute with the same sign to $\Delta S$.
Hence the absolute value of the action difference can never be small and
the associated contributions vanish after averaging over the energy.
The simplest examples are thus \be
\widetilde\R_2(\ep_1,\ep_2)=\widetilde\R_{1,1}(\ep_1,\ep_2),\ee and \begin{align}
\widetilde\R_3(\ep_1,\ep_2,\ep_3)&=\widetilde\R_{1,2}(\ep_1,\ep_2,\ep_3)+\widetilde\R_{2,1}(\ep_1,\ep_2,\ep_3)\nonumber\\&=2{\rm
Re}\widetilde\R_{1,2}(\ep_1,\ep_2,\ep_3).\label{R3real}\end{align}

As already argued, averaging over the energy annihilates every summand apart from those
with small action differences, i.e. almost identical cumulative actions of $P$ and $Q$.
The simplest and important case of identical orbits will be discussed in the next
subsection. The general mechanism behind non-trivial action correlations has been
analyzed extensively in previous works \cite{sieber,sieber2,R2a,R2b,R2c,longa,longb}.
This is that each $q$-orbit must follow closely (up to time reversal) a certain $p$-orbit
for a period of time. However it can switch to be close to a different $p$-orbit (or a different part of the same $p$-orbit) in  what is called an
\emph{encounter}. An $\ell$-encounter is a region where $\ell$ stretches of $p$-orbits
run nearly parallel (i.e. close in phase space) or anti-parallel (i.e. mutually time-reversed).
The $q$-orbits then differ from the $p$-orbits by differently connecting the endpoints of these encounter stretches. (See Figures 1
and 2 for illustrations.) 
Outside the encounters the $q$-orbits are nearly equal to the $p$-orbits.
In the following the orbit parts outside the encounters are referred to as links.

\subsection{Diagonal Approximations}

The simplest contribution to the average in (\ref{RJK}) comes from the so-called diagonal
approximation. For systems without TRS, this approximation accounts for the
case that all orbits are pairwise equal, i.e. $P=Q$ (for $n=2$ this was
considered in \cite{hoda,berry}). This situation is only possible for $J=K$ and hence
even $n$.

We start by considering two orbits $p$ and $q$
that are identical apart from having two slightly different energies
$E_p=E+\ep/\bar\rho$ and $E_q=E+\eta/\bar\rho$.
We can neglect differences between the periods  $T_p$ and $T_q,$ and  between the stability factors $F_p$ and $F_q$.
For the actions (whose difference will be divided by $\hbar\to0$) 
we a have to be more careful and Taylor expand to linear order, using\be \frac{\partial}{\partial\epsilon}
S_p(E+\epsilon/\bar{\rho})=\frac{T_p}{\bar{\rho}}.
\label{deriv}\ee  
This leads to
$S(E_p)\approx S(E)+\ep T_p/\bar\rho$ and thus $\Delta S\approx(\ep-\eta)T_p/\bar\rho$. The contribution of two identical orbits to $R_{J,K}$ can now be evaluated using
the Hannay-Ozorio de Almeida sum
rule \cite{hoda}. In the notation of \cite{longb} this rule can be written as
\be\label{hodaeq}\sum_p |F_p|^2 f(T_p)\approx\int_0^\infty \frac{dT}{T}f(T)
\ee
where $f$ represents any function of an orbit that depends only on its period.
 
The contribution of identical orbits can now be evaluated as\begin{align}\label{diag} \frac{1}{T_H^2}\sum_p |F_p|^2 T_p^2e^{i(\ep-\eta)T_p/\bar\rho\hbar}&=\frac{1}{T_H^2}\int_0^\infty
dTTe^{i(\ep-\eta)T/\bar\rho\hbar}\nonumber\\&=-\frac{1 }{4\pi^2(\ep-\eta)^2},\end{align}
where the integral is regularized by adding a small positive imaginary part to $\ep$. 
For sets of orbits $P$ and $Q$ that coincide pairwise a factor of this type is obtained for each pair of orbits.
For time-reversal invariant systems the result of (\ref{diag}) must be multiplied
by 2 to account for pairs of mutually time-reversed orbits leading to
\be
\label{diag_orth}
-\frac{1 }{2\pi^2(\ep-\eta)^2}.
\ee
 
For $n>3$ there is also the possibility of \emph{partial diagonal approximations}, in
which only a few orbits coincide, i.e. we may have $P'=Q'$ with $P'\subsetneq P$ and
$Q'\subsetneq Q$. The simplest such example is at $n=4$ and $J=K=2$
 meaning that $P$ consists of two orbits $p_1,p_2$
 and $Q$ consists of two orbits $q_1,q_2$.
  In this case $R_{2,2}(\epsilon,\eta)$ contains a contribution, denoted
  by $R_{2,2}^{(0)}$ from $P$, $Q$ in which no two orbits are the same. In
  addition there is a contribution from $P,Q$ that share one orbit,
\begin{eqnarray}\label{partial}
-\frac{1}{4\pi^2}\Big(&&\frac{R^{(0)}_{1,1}(\ep_2,\eta_2)}{(\ep_1-\eta_1)^2}+
\frac{R_{1,1}^{(0)}(\ep_2,\eta_1)}{(\ep_1-\eta_2)^2}\nonumber\\ 
+&&\frac{R^{(0)}_{1,1}(\ep_1,\eta_2)}{(\ep_2-\eta_1)^2}+
\frac{R_{1,1}^{(0)}(\ep_1,\eta_1)}{(\ep_2-\eta_2)^2}\Big),\end{eqnarray}
obtained by multiplying diagonal terms $-\frac{1}{4\pi^2(\ep_j-\eta_k)^2}$ associated with two coinciding orbits   with off-diagonal
contributions $R_{1,1}^{(0)}$ accounting for the two remaining orbits.
For each contribution, the off-diagonal factor must involve the energy increments not appearing in the diagonal factor. A final contribution arises from the case that all orbits coincide pairwise.

\section{Semiclassical diagrammatics}

Using once more (\ref{deriv})  we can rewrite (\ref{RJK}) as \be\label{R2A}
R_{J,K}(\epsilon,\eta)=\frac{(-1)^K}{(2\pi i)^n} \prod_{j=1}^J\frac{\partial}{\partial
\epsilon_j} \prod_{k=1}^K\frac{\partial}{\partial \eta_k}\mathcal{A}_{J,K}(\ep,\eta), \ee
where $\mathcal{A}_{J,K}(\ep,\eta)$ stands for the energy average \be
\mathcal{A}_{J,K}(\ep,\eta)=\left\langle\sum_{P,Q} F_P F_Q^* e^{i\Delta
S/\hbar}\right\rangle.\ee

In this Section we consider the situation when there are no coinciding orbits, i.e. we
treat the quantity $R_{J,K}^{(0)}(\epsilon,\eta)$, and the analogously defined
$\mathcal{A}^{(0)}_{J,K}(\ep,\eta)$.

The quantity $\Delta S$ has two contributions. The first one we have already met: the
orbits have slightly different energies. This contribution can be approximated by \be
\sum_{j=1}^JT_{p_j}\epsilon_j/\bar\rho-\sum_{k=1}^K T_{q_k}\eta_k/\bar\rho.\ee The second
contribution depends on the separation between the orbit stretches taking part in the
encounters. For an encounter $e$ involving $\ell_e$ stretches, there are $\ell_e-1$
relative separations. For a system with two degrees of freedom, each of these can be
decomposed into two components, pointing along the direction of the stable and unstable
manifolds, denoted by $s_{e,m}$ and $u_{e,m}$. The second contribution to the action
difference is then obtained as $\sum_e\sum_{m=1}^{\ell_e-1}s_{e,m}u_{e,m}$.
For details of the derivation, and the definition of $s_{e,m}$ and $u_{e,m}$,
we refer to \cite{R2c} (based on \cite{TR,Sp} for $l_e=2$).

In the semiclassical limit the correlation functions are dominated by pairs
of orbits with action differences at most of the order $\hbar$. As a consequence
the separations in the relevant encounters are very small and the $p$ and
$q$-orbits are very close inside the encounters
as well as outside. As a consequence we can approximate $F_Q\approx F_P$
and hence $F_PF_Q^*\approx |F_P|^2$.
  
The sum over the set of correlated trajectories, $Q$, can now be replaced by an integral over
an ergodic probability density, $w(s,u)$, determining the likelihood of encounters with
given action difference. For the simplest correlation function, requiring only two
orbits, this was done in detail in \cite{R2c} (see the appendix of \cite{transport} as
well as \cite{handbook} for the formulation using derivatives as in Eq. (\ref{R2A})) and yields (in our present notation)
\be\label{R2struc} \sum_{\rm struc}c_{\rm struc} \frac{(-1)^V}{(-2\pi
i(\epsilon-\eta))^{L-V}},\ee as a semiclassical approximation to
$\mathcal{A}_{1,1}^{(0)}(\epsilon,\eta)$. Here $V$ is the total number of encounters and
$L$ is the total number of `links', trajectory pieces connecting encounters.

The summation in (\ref{R2struc}) is over the possible {topological} \emph{structures}
that the encounters can produce. These structures are characterized by the number of
encounters, the number of stretches belonging to each encounter, the way the encounter
stretches are distributed among the orbits, and their ordering along the orbits. For
time-reversal invariant systems they also depend on whether the encounter stretches point
in the same direction or are time reversed. We will later see that the structures can be
conveniently described in terms of permutations. The factor $c_{\rm struc}$ avoids
overcounting due to some subtleties of the definition of structures that will be
discussed at a later stage.

In the present situation of more general values of $J$ and $K$, an analogous calculation
can be performed. Previous works \cite{longb} have established that
for orbits of periods $T_{p_j}$
a suitable probability density to find encounters associated to a given structure
and with given stable and unstable separations is given by\be
w(s,u)=\frac{\prod_j T_{p_j}\int dt}{\prod_e \Omega^{\ell_{e}-1}t_{e}}.\ee Here $\Omega$
is the volume of the energy shell, \be t_{e}=\frac{1}{\lambda}\ln\frac{c^2}{\min_m
|s_{e,m}|\min_n|u_{e,n}|}\ee is the duration of the encounter $e$, $\lambda$ is the Lyapunov
exponent of the system, and $c$ is a constant that will be irrelevant for the final
result. $\int dt$ denotes a multiple integral over all times at which the orbits in $Q$
traverse the encounters, reckoned from a reference traversal. Now the sum over $Q$ for
each given $P$ can be replaced by an integral over the density $w(s,u)$. Finally, the remaining sum
over $P$ is evaluated using the Hannay-Ozorio de Almeida sum rule (\ref{hodaeq}).
The  contribution to the sum $\left\langle\sum_{P,Q} F_P F_Q^* e^{i\Delta S/\hbar}\right\rangle$  from each structure is
\begin{widetext}
\begin{equation}
\int_0^\infty\frac{dT_{p_1}}{T_{p_1}}...
\int_0^\infty\frac{dT_{p_J}}{T_{p_J}}\int \prod_e\prod_{m=1}^{\ell_e-1}ds_{e,m}du_{e,m} w(s,u)\exp\left(\frac{i}{\hbar}\left[
\sum_{j=1}^J\frac{T_{p_j}\epsilon_j}{\bar\rho}-\sum_{k=1}^K \frac{T_{q_k}\eta_k}{\bar\rho}+ \sum_e\sum_{m=1}^{\ell_e-1}s_{e,m}u_{e,m}\right]\right)\nonumber.
\end{equation}
\end{widetext}

Luckily,
this expression factorizes
nicely into contributions associated to encounters and links.  The divisors $T_{p_j}$ cancel with the corresponding factors in $w(s,u)$. Afterwards the integrals over orbit periods and the integrals over encounter traversal times
in $w(s,u)$ can be transformed into integrals over link durations, which we again denote by $t$. The only
contribution of a {\it link} to the action difference is due to the different energy
increments. A link that belongs to the orbit $p_j$ of $P$ and the orbit $q_k$ of $Q$
gives a contribution $t(\ep_j-\eta_k)/\bar\rho$ to $\Delta S$. If we integrate and
incorporate as a factor the inverse of the Heisenberg time, we obtain \be\label{link} T_H^{-1}\int
dte^{it(\ep_j-\eta_k)/(\hbar\bar\rho)}=-\frac{1}{2\pi i (\epsilon_j-\eta_k)}.\ee

The contribution of each {\it encounter} to
$\sum_{j=1}^JT_{p_j}\epsilon_j/\bar\rho-\sum_{k=1}^K T_{q_k}\eta_k/\bar\rho$ depends on
the numbers of stretches it involves that form part of the different orbits. We assume
that the encounter involves $\ell_j$ stretches of each of the orbits $p_j$, and after
changing connections it involves $\widetilde \ell_k$ stretches of each of the orbits
$q_k$ (with $\sum_j \ell_j=\sum_k \widetilde \ell_k=\ell)$. Then its contribution to the
above sum can be written as $(\sum_j\ell_j\epsilon_j-\sum_k\widetilde
\ell_k\eta_k)t_{e}/\bar\rho$. 

To obtain the contribution to $\mathcal{A}^{(0)}_{J,K}(\ep,\eta)$  must also take into account the remaining part of the
action difference given by $\sum_e\sum_{m=1}^{\ell_e-1}s_{e,m}u_{e,m}$, the density $w(s,u)$ as well as  factors $T_H^\ell$ that will
altogether compensate the factor inserted in the link contribution. When all of this is
taken into account we arrive at
a factor\begin{widetext}\be \label{encount}T_H^{\ell}\int d^{\ell-1}s\,d^{\ell-1}u\frac{1}{\Omega^{\ell-1}t_{{e}}}
\exp\Big[{i\sum_m u_{ m}s_{m }/\hbar}+\Big(\sum_j\ell_j\epsilon_j-\sum_k\widetilde
\ell_k\eta_k\Big) t_{e}/(\bar\rho\hbar)\Big] \approx 2\pi
i\Big(\sum_j\ell_j\epsilon_j-\sum_k\widetilde \ell_k\eta _k\Big)\ee
\end{widetext}
arising from every encounter (where we suppressed the subscripts $e$ of $\ell$,
$s$, and $u$).

As in previous works, the above integral is computed by expanding the $t_{e}$-dependent
part of the exponential into a power series. Contributions relevant in the semiclassical
limit arise only from the linear term in that expansion. They can be
calculated by noting that the factor $t_{e}$ is canceled by the divisor $t_{e}$ in front
of the exponential, and furthermore using  $\int du ds e^{ius/\hbar}=2\pi\hbar$ and
$\bar\rho=\frac{\Omega}{(2\pi\hbar)^2}$. By contrast, the integral of the leading term in
the expansion leads to a result that vanishes after averaging over the energy, and the
higher-order terms are negligible in the semiclassical limit.

The factor (\ref{encount}) associated with an encounter can also be
written in a slightly different but equivalent and technically advantageous
way. To do so we single out one of the $\ell$ points where the orbit enters this
encounter as the `first'. Say this point (and the preceding link) belongs to a certain
$p_j$ and a certain $q_k$. If we assign to the encounter a factor 
\be\label{enc}
2\pi
i(\epsilon_j-\eta_k)
\ee
and sum over all choices of first stretches, we obtain the same
result as (\ref{encount}), since in this sum each index $j$ arises $\ell_j$ times and
each index $k$ arises $\widetilde \ell_k$ times. Hence, these two ways of dealing with
the encounter factor are equivalent. However, this second method (assigning to the
encounter a quantity which depends on its `first' stretch and then summing over all
possible first stretches) is more convenient, because the encounter factors become
inverse to link factors and just compensate the links that are the first to arrive at the
encounter.

In line with the above discussion, let $M_{jk}$ be the number of times orbits $p_j$ and
$q_k$ run together in a link but are not the first ones to arrive at an encounter. Then each of these $M_{jk}$ links gives a factor $-\frac{1}{2\pi i (\epsilon_j-\eta_k)}$,
see (\ref{link}). The encounter contributions (\ref{enc}) are cancelled apart from a factor $-1$ per encounter.
Altogether, summation over structures thus gives 
the   
semiclassical approximation   
\begin{align}\label{TJK} \mathcal{A}_{J,K}^{(0)}(\ep,\eta)&=\sum_{\rm struc}\prod_{jk}c_{\rm
struc} \frac{(-1)^V}{(-2\pi i(\epsilon_j-\eta_k))^{M_{jk}}}\\&=\sum_{\rm
struc}\frac{(-1)^L}{(2i\pi)^{L-V}}\prod_{jk} c_{\rm struc}\, z_{jk}^{M_{jk}}.\label{sum}\end{align}
Here we have denoted the number of encounters by $V$, the number of links by $L$, and used that $\sum_{jk}M_{jk}=L-V$.
We have also introduced 
\be z_{jk}=\frac{1}{\ep_j-\eta_k} \label{xdef}.\ee 
The quantity $c_{\rm struc}$ will be
discussed properly in the next Sections; in particular, it depends on whether
the system
is time-reversal invariant or not. 

Finally $R^{(0)}_{J,K}(\ep,\eta)$
can be accessed from these results by performing the derivatives given in (\ref{R2A}). If we absorb these derivatives in 
 \be\label{defderiv} D_{J,K}=\frac{(-1)^K}{(2\pi i)^n}
\prod_{j=1}^J\frac{\partial}{\partial\ep_j}\prod_{k=1}^{K}\frac{\partial}{\partial\eta_k},\ee
we can write our semiclassical result for $R^{(0)}_{J,K}(\ep,\eta)$ as 
\be\label{Rnsemi} D_{J,K}\Big[\mathcal{A}_{J,K}^{(0)}\Big].\ee

\section{Structures and permutations, for broken TRS}

In order to formalize the concept of a structure, it is useful to introduce permutations
\cite{R2a,R2b,R2c,combinat1,combinat2,combinat3,combinat4}. We number the encounter
stretches in $P$, from $1$ to $L$, in such a way that inside the encounters the
trajectories in $Q$ go from the beginning of stretch $i$ to the end of stretch $i+1$.   
The beginning of the final stretch of each encounter is then connected to the end of the initial one. This mapping can be expressed by a permutation $\sigma$. It is useful to reduce the freedom in
ordering the encounters by requiring that longer ones come first. In this way, we
associate with the set of encounters a permutation $\sigma$ whose cycles are
\be\label{sigmas}\sigma=(12\cdots \ell_1)(\ell_1+1\cdots \ell_1+\ell_2)\cdots,\ee where
$\ell_1\ge \ell_2\ge\cdots$ are the encounter sizes.

On the other hand, we associate with $P$ the permutation $\pi$ which takes $i$ to $j$ if
there is a link from the end of stretch $i$ to the beginning of stretch $j$. Successive application of $\pi$ hence yields the
encounter stretches included in each of the orbits of $P$.
Thus, clearly the
cycles of $\pi$ correspond to the orbits in $P$. 

Finally, we associate with trajectories
$Q$ the permutation $\rho$ which is the product $\rho=\pi\sigma$. (Here products of permutations are defined such that when they are applied to a number the right-most factor is applied first.)  

Applying this product to the index of the start point of an encounter stretch leads first
to the end point it is connected to along $Q,$ and then to the start point of the stretch
following the next link. Repeated application thus enumerates the start points in the
order they are visited by $Q$. The cycles of $\rho$ are therefore in one-to-one relation to
the orbits in $Q$.

Notice that there is still freedom in the relative order of encounters of the same size.
Also, we must choose the first stretch in each encounter, which will be labelled by the
smallest number. These different choices do not alter the permutation $\sigma$, but may
alter $\pi$ and $\rho$. We will take into account all possible such choices. It is
crucial to make sure that this does not lead to overcounting of the contributions to the
correlation function. If we denote the number of encounters with $\ell$ stretches by $v_\ell$,
there are $v_\ell!$ ways of ordering these encounters. To avoid overcounting, we thus have
to divide out $v_\ell!.$ (Note that we could equivalently have left the ordering of
encounters completely unspecified, and then divided by the factorial $V!$ of the overall
number of encounters.) As already described earlier, we take into account all choices of
a first stretch inside each encounter. However we have already modified the contribution
of the encounter such that summation over all these choices leads to the correct
encounter factor derived from semiclassics. Hence no further corrections are necessary
and the factor $c_{\rm struc}$ in (\ref{Rnsemi}) has to be chosen as \be\label{c_unit} c_{\rm
struc}=\frac{1}{\prod_\ell v_\ell!}.\ee

\begin{figure}[t]
\includegraphics[scale=0.5,clip]{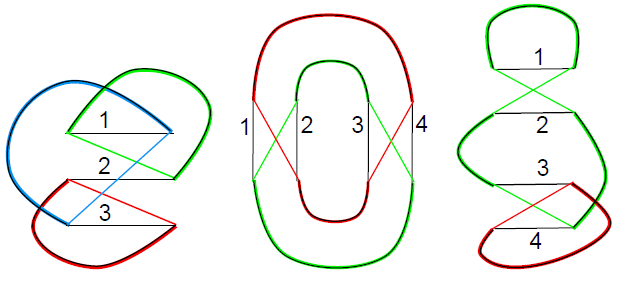}
\caption{(color online) Correlated periodic orbits. Encounters are grossly exaggerated.
Orbits in set $P$ are depicted in black lines, orbits in $Q$ in colored/grey lines. Left
diagram has $J=1$, $K=3$; only one structure is possible. Middle diagram has $J=K=2$; two
structures are possible, one of them is shown. Right diagram also has $J=K=2$; only one
structure.} \label{Fig1}
\end{figure}

The example in Figure 1 can help visualize  the permutations we have introduced. In the
first diagram, the encounter permutation is $\sigma=(123)$, the permutation associated
with the black orbit (the only member of $P$) is $\pi=(132)$, and the permutation associated with the three colored
(grey) orbits (forming $Q$) is $\rho=\pi\sigma=(1)(2)(3)$. This diagram has only one possible choice of
permutations, because changing the first stretch does not change either of the
permutations. 

In the second diagram, with the labels shown in the figure, we have
$\sigma=(12)(34)$, $\pi=(14)(23)$ and $\rho=(13)(24)$. However, there is an equivalent
choice of the permutations $\pi$ and $\sigma$: $\pi=(13)(24)$ and $\rho=(14)(23)$. Here the labels of the stretches 3 and 4 are exchanged. This leads to a different structure that has to be taken into account separately. 

 The last diagram also has
$\sigma=(12)(34)$, but it admits four different choices of the remaining permutations:
$\pi=(123)(4)$, $\rho=(134)(2)$; $\pi=(124)(3)$, $\rho=(143)(2)$; $\pi=(132)(4)$,
$\rho=(1)(234)$; $\pi=(142)(3)$, $\rho=(1)(243)$.

The permutation $\sigma$ is fixed once the encounter sizes are known, and the equation
$\sigma=\pi^{-1}\rho$ can be seen as a factorization of $\sigma$. As in the final two examples above, different choices of
first stretch inside each encounter produce different structures if they lead to
different factorizations. On the other hand, if we   exchange e.g.  $1\leftrightarrow
3$,
$2\leftrightarrow 4$ in the middle   diagram of Figure 1 this simply  exchanges the encounters and does not lead to a different factorization or  structure.

The calculation of $\mathcal{A}_{J,K}^{(0)}$ requires factorizations of specific permutations,
of the kind seen in Eq.(\ref{sigmas}). In these factorizations, the first factor, $\pi^{-1}$,
must have $J$ cycles, while the second factor, $\rho$, must have $K$ cycles,
respectively corresponding to the orbits of $P$ and $Q$.
The
factorization must take place in the group of permutations of $L$ symbols (the number of
links), and $\sigma$ must have $V$ cycles (the number of encounters). Notice that
$\sigma$ cannot have fixed points, since encounters have size at least $2$. Therefore,
for a given order in perturbation theory, i.e. for a fixed value of $L-V$, there exist
only a finite number of factorizations.

Each structure is then characterized by (i) a choice of the permutations $\sigma$, $\pi$
and $\rho$ subject to these requirements, but crucially also (ii) one choice of assigning
the $J$ ($K$) cycles of $\pi$ ($\rho$) to the $J$ ($K$) orbits in $P$ ($Q$).

\section{Structures and permutations, for preserved TRS}

The structures arising for time-reversal invariant systems can also be described in terms
of permutations. In this case we also have to account for the different directions of
motion. The resulting permutations will describe the connections of the orbits in $P$ and
$Q$ as well as their time-reversed versions.

To define the encounter permutations $\sigma$, we arbitrarily single out one preferred
direction of motion inside each encounter. We then label the stretches in that direction
of motion (belonging either to the orbits of $P$ or their time-reversed versions) by
consecutive integers. The time-reversed version of each stretch, going opposite to the
preferred direction, is indicated by the same integer but with an overbar. So the start of stretch $\overline a$ has the same position (but opposite sense of motion) as the end of stretch $a$, and the end of stretch $\overline a$ coincides with the beginning of stretch $a$ (again up to sense of motion).

The permutation $\sigma$ maps the start point of each stretch to the endpoint it is connected
to within $Q$, i.e., after switching connections.

As before, the connections of the stretches in the preferred direction are indicated by
cycles $(12\cdots\ell_1)(\ell_1+1\cdots\ell_1+\ell_2)\cdots$. If a stretch connects the
start of $a$
 to the end of $b$, its time-reversed will connect the
start of the time-reversed of $b$, denoted by $\bar b$, to the end of $\bar a$. Hence
each cycle is accompanied by one with all elements barred and reversed in order, leading
to
\begin{equation}\begin{alignedat}{2}
\sigma=(12\cdots\ell_1)(\ell_1+1\cdots\ell_1+\ell_2)\cdots\\\cdots
(\overline{\ell_1+\ell_2}\cdots\overline{\ell_1+1})(\bar \ell_1\cdots \bar 2\bar 1).
\end{alignedat}\end{equation}

The permutation $\pi$ determines how the end points of encounters are connected to start
points through links. $\pi$ may map indices with bars to indices without or the other way
around; this happens for example if the link returns to the same encounter but with
opposite sense of motion, or if it leads to a different encounter but the preferred
directions of motion in these encounters are not aligned. Similarly as above, if $\pi$
maps a given end point to a given start, $a\to b$, it induces the opposite mapping
between the time-reversed versions of these points, $\bar b\to \bar a$. 
(Note that we define $\overline{\overline a}=a$.)
Hence a cycle of
the form, say, $(ab\bar c )$ will be accompanied by a second cycle of the form $(c\bar
b\bar a)$. This is the only restriction on the form of $\pi$.

As inside each encounter of $P$ (or its time-reversed) every start point is  connected to
the end point with the same index, application of $\pi$ enumerates the start points in the
order of traversal by $P$ and its time reversed. Hence every cycle of $\pi$ will
correspond to a periodic orbit in $P$ or its time reversed. Pairs of cycles like $(ab\bar
c)$ and $(c\bar b\bar a)$ describe mutually time-reversed orbits.

\begin{figure}[b]\includegraphics[scale=0.5,clip]{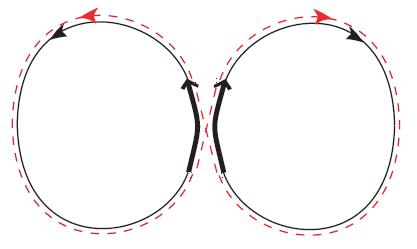}\caption{ (color online) Diagram giving the leading contribution to $\R_3$ (encounters are
grossly
exaggerated). There is one orbit, depicted by a dashed line, correlated with
two others, depicted by a full line. (Figure from \cite{longb}.)} \label{butterfly}
\end{figure}

In analogy to  systems without time-reversal invariance, the product $\rho=\pi\sigma$
maps the start point of each encounter stretch to the start point of the stretch
following along $Q$ or its time-reversed. The cycles of $\rho$ come in pairs where one
cycle describes an orbit in $Q$ and the other one describes its time-reversed version.
(The relation between the cycles of $\rho$ is more complex than for $\sigma$ and
$\pi$, and it does not lead to any further constraints on the factorizations used.)

\begin{figure*}[t]
\includegraphics[scale=0.4,clip]{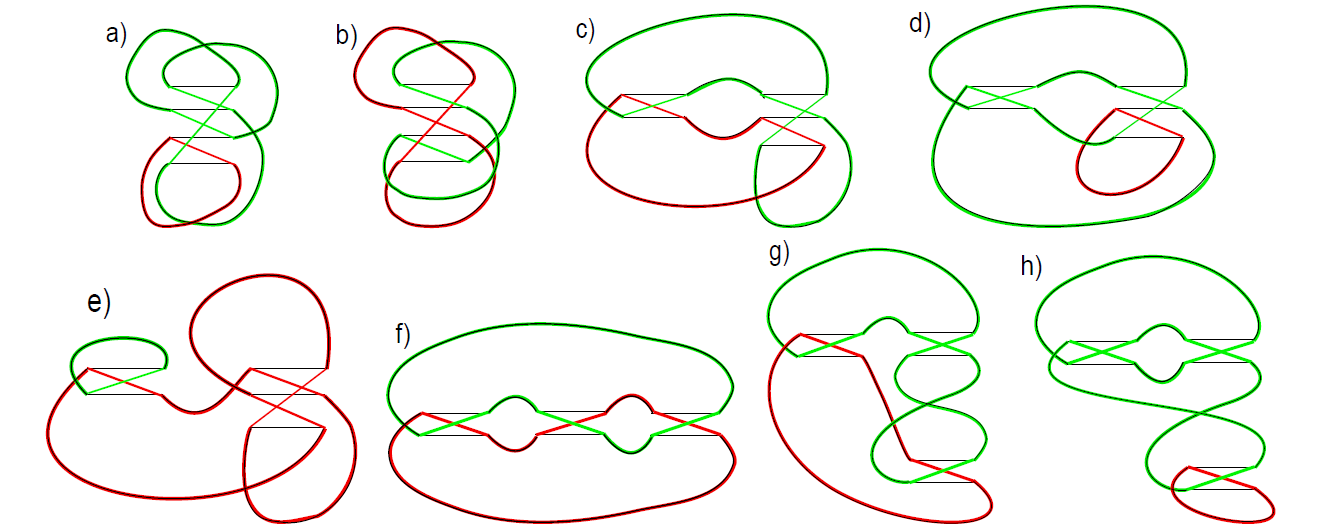}
\caption{ (color online) Diagrams depicting correlated periodic orbits contributing to
the third order perturbation theory calculation of $\R_3$ (encounters are grossly
exaggerated). In every case we have one orbit, in black, correlated with two others, in
red and green (grey). } \label{Fig2}
\end{figure*}

In the permutations thus defined, the orbits and their time-reversals are treated on
equal footing. For each pair of time-reversed orbits described by $\pi$ there are two
possible choices for the orbit to be included in $P$. Similarly for each pair of orbits
described by $\rho$ there are two choices for the orbit to be included in $Q$. Hence, if
we want to sum over all choices for $P$ and $Q$, we have to include a factor $2^{J+K}$,
where $J$ is the number of orbits in $P$ (half the number of cycles in $\pi$) and $K$ is
the number of orbits in $Q$ (half the number of cycles in $\rho$). On the other hand, for
every choice of $P,Q$ we take into account all $2^V$ ways of fixing preferred directions
inside the encounters. To avoid overcounting, we have to divide out $2^V$. Together with
the division by $\prod_\ell v_\ell!$ explained earlier, we thus need a factor
\begin{equation}
c_{\rm struc}=\frac{2^{J+K-V}}{\prod_\ell v_\ell!}\label{c_orth}.
\end{equation}

\section{Leading orders, broken TRS}

Our diagrammatic rule, Eq.(\ref{Rnsemi}), can be used for both unitary and orthogonal
universality classes. By constructing the simplest diagrams explicitly (or finding the
relevant factorizations), it is possible to obtain the first orders in perturbation
theory.

For the unitary class, corresponding to broken time-reversal symmetry, the leading order
approximation to $\R_n$, stemming from the diagonal approximation, has been obtained by
Nagao and M\"uller \cite{taro}. It turns out that this `approximation' is in exact agreement with the
prediction from random matrix theory. Here we show that the first few perturbative
corrections indeed vanish for the simplest functions.

\subsection{$2$-point function}

The $2$-point function has already been the subject of many papers, the closest one to
the present approach being \cite{R2c}. We discuss it again briefly.

The difference between the present approach and the one in \cite{R2c} is in the concept
of structure and in the way the encounter stretches are numbered. Since there is only one
orbit in the set $P$, it made sense in \cite{R2c} to just number the stretches in the
order they were visited by that orbit (the same convention was also used in
\cite{combinat3}). As a result, this set was always represented by the permutation
$P_{\rm loop}=(12\cdots L)$ and structures were identified with factorizations of this
permutation in which one of the factors was also a single-cycle permutation, representing
the single-orbit set $Q$.

In this work, we are fixing the encounter permutation to be of the form of
Eq. (\ref{sigmas}). Therefore, neither of the single-cycle permutations representing $P$
or $Q$ need be given by $(12\cdots L)$. Instead, we define structures in terms of
factorizations of $\sigma$ into single-cycle factors.

In \cite{R2c} it was shown that all off-diagonal contributions to the 2-point
function cancel for systems without TRS. We want to briefly discuss how this
plays out in the leading off-diagonal order with our present conventions.
As in \cite{R2a} this order is determined by a diagram involving one 3-encounter
and a diagram involving two 2-encounters. The former diagram has one structure
described by the permutations $\rho=(123)$, $\pi=(123)$, $\rho=(132)$, and the
latter diagram has two structures with 
$\sigma=(12)(34)$, $\pi=(1423)$, $\rho=(1324)$ and $\sigma=(12)(34)$, $\pi=(1324)$,
$\rho=(1423)$. Their contributions to $\T_{11}$ cancel as the factor $\frac{(-1)^L}{\prod_\ell
v_\ell!}$ from Eqs. (\ref{sum}) and (\ref{c_unit}) is equal to $-1$ for the
first diagram and equal to $\frac{1}{2}$ for both structures of the second
diagram.

\subsection{$3$-point function}

The leading-order correction to the $3$-point correlation function $\R_3$ consists in a
single diagram, where $P$ contains a single orbit with a single $2$-encounter, correlated
with two orbits in the set $Q$, see Figure \ref{butterfly}. This
possibility has a single structure, associated with $\sigma=(12)$, $\pi=(12)$,
$\rho=(1)(2)$, i.e. the factorization $(12)=
(12)\cdot(1)(2)$. One of the orbits in $Q$ is always the first to arrive at the encounter, so the
semiclassical contribution is proportional to either $z_{11}=(\ep_1-\eta_1)^{-1}$ or
$z_{12}=(\ep_1-\eta_2)^{-1}$. In any case, it depends only on two variables. When we act
with the operator $D_{12}\propto\partial ^3/\partial \ep_1\partial\eta_1\partial\eta_2$,
as required by (\ref{Rnsemi}), the final result vanishes.
The same happens if the single orbit is taken as the only element of $Q$ and the two orbits are included in $P$.

This mechanism is quite general. According to (\ref{Rnsemi}), we must take derivatives
with respect to all energies. But when a structure contains an orbit which participates
in only one encounter, and it is the first one to arrive at that encounter, the quantity
$\prod_{jk} z_{jk}^{M_{jk}}$ is independent of the energy of that orbit, and the
derivative vanishes.

\begin{table}
\begin{center}
\begin{tabular}{|c|c|c|c|c|}\hline
& $\sigma$ & $\pi$ & $\rho=\pi\sigma$ & mult.  \\ 
\hline 
(a) & (1234) & (1243) & (142)(3) & 3\\
 \hhline{~----}
 & \gr (1234) & \gr (1342) & \gr (1)(243) & \gr 1\\
 \hline 
(b) & (1234) & (1234) & (13)(24) & 1\\
 \hline 
(c) & (123)(45) & (12534) & (15)(243) & 5\\
 \hhline{~----}
 & \gr (123)(45) & \gr (12435) & \gr (14)(253) & \gr 1\\
 \hline 
(d) & (123)(45) & (13425) & (1524)(3) & 4\\
 \hhline{~----}
 & \gr (123)(45) & \gr (14352) & \gr (1)(2534) & \gr 2\\
 \hline 
(e) & (123)(45) & (12345) & (1324)(5) & 3\\
 \hhline{~----}
 & \gr (123)(45) & \gr (12354) & \gr (1325)(4) & \gr 3\\
 \hline 
 
 (f) & (12)(34)(56) & (146235) & (136)(245) & 6\\
 \hhline{~----}
 & \gr (12)(34)(56) & \gr (145236) & \gr (135)(246) & \gr 2\\
 \hline 
 
(g) & (12)(34)(56) & (162453) & (14)(2635) & 18\\
 \hhline{~----}
 & \gr (12)(34)(56) & \gr (162354) & \gr (13)(2645) & \gr 6\\
 \hline 
(h) & (12)(34)(56) & (162345) & (13526)(4) & 24\\
 \hhline{~----}
 & \gr (12)(34)(56) & \gr (162435) & \gr (14526)(3) & \gr 24\\
 \hline
\end{tabular}
\end{center}
\caption{Structures associated to the diagrams of Figure \ref{Fig2}. Encounters
are represented by a permutation $\sigma$, the black orbit by $\pi$ and the
other orbits by $\rho$. The structures marked in grey give contributions
that vanish after taking derivatives.
\label{table124GUE}}
\end{table}

The above contribution involved $L-V=1$, and 
a simple argument involving the parities of permutations shows 
that for systems without TRS every second value of $L-V$ does not have any associated diagrams. (The parity of a permutation with $L$ elements and $C$ cycles is given by $(-1)^{L-C}$. Hence
the parities of $\pi$, $\rho$ and $\sigma$ are given by
$(-1)^{L-J}$, $(-1)^{L-K}$ and $(-1)^{L-V}$; using $\rho=\pi\sigma$ one can then show that $L-V$ must be even if $J-K$ is even and odd otherwise.)
Hence the next correction to  $\R_3$ arises from $L-V=3$. In Figure 3, we
sketch the correlated sets of periodic orbits that are important in this case. There are
two diagrams with a single $4$-encounter. One of them has four possible choices of
permutations, while the other admits a single choice. There are three diagrams
(altogether 18 choices of permutations) with a $2$-encounter and a $3$-encounter, and
three other diagrams (altogether 80 choices) with three $2$-encounters.

In Table \ref{table124GUE}
 we present all the permutations/structures that the diagrams in
Figure \ref{Fig2} may have. 
The diagrams allow for different choices of permutations that can give different
contributions. In the table we are displaying one representative choice for each contribution, and we give the number of overall choices with the same contribution in the final column. For example the additional choices of permutations
for diagram (a) are $\pi=(1324)$, $\rho=(143)(2)$ and $\pi=(1423)$, $\rho=(132)(4)$. The contributions marked in grey are proportional to $z_{11}^3$ and hence vanish after taking derivatives. Notice how in all these structures the permutation $\rho$ has a cycle involving only numbers which begin a cycle in $\sigma$. 

The contributions of the remaining structures in Table~\ref{table124GUE} to $\T_{1,2}$ are proportional to
$z_{11}^2z_{12}$ or to $z_{11}z_{12}^2$, depending on how the energy increments $\eta_1$
and $\eta_2$ are assigned to the orbits. The result after symmetrization is proportional
to ${\cal N}(z_{11}^2z_{12}+z_{11}z_{12}^2)$. Here $\mathcal{N}$ is the sum over
structures taking into account the multiplicities in the table as well as the weight
 $\frac{(-1)^L}{\prod_\ell v_\ell!}$ arising from (\ref{sum}) and (\ref{c_unit}), with $V$ the total number of permuted elements.
The latter weight gives a minus sign for diagrams (c) and (e), and a factor $\frac{1}{6}$ for diagrams (f) to (h).
It is easy to read off Table I that $\mathcal{N}$ is   equal to
$\mathcal{N}=3+1-5-4-3+1+3+4=0,$ as expected.

\subsection{4-point function}

The leading order correction to the $4$-point correlation function has $L-V=2$. Let us
consider separately the quantities $\T_{1,3}$ and $\T_{2,2}$.

The former has two different contributing diagrams. One diagram has a single
$3$-encounter and only one possible structure; this is the first diagram shown in Figure
1. The other diagram has two $2$-encounters, hence $\sigma=(12)(34)$, and three choices
of permutations:
 $\pi=(1234)$, $\rho=(13)(2)(4)$; $\pi=(1342)$, $\rho=(1)(23)(4)$;
$\pi=(1432)$, $\rho=(1)(24)(3)$. All of these  have vanishing contributions.
These arise from derivatives of terms proportional to $z_{1k}z_{1k'}$ (where $k,k'=1,2,3$), which must vanish as at least one of the choices $1,2,3$ for the second index is absent.

\begin{table} 
\begin{center}
\begin{tabular}{|c|c|c|c|c|}\hline
& $\sigma$ & $\pi$ & $\rho=\pi\sigma$ & mult.  \\ 
\hline 
(a) & (123) & (13)(2) & (12)(3) & 1\\
 \hhline{~----}
 & \gr (123) & \gr (12)(3) & \gr (1)(23) & \gr 1\\
 \hhline{~----}
 & \gr (123) & \gr (1)(23) & \gr (13)(2) & \gr 1\\
 \hline 
(b) & (12)(34) & (134)(2) & (123)(4) & 2\\
 \hhline{~----}
 & \gr (12)(34) & \gr (143)(2) & \gr (124)(3) & \gr 2\\
 \hhline{~----}
 & \gr (12)(34) & \gr (1)(234) & \gr (132)(4) & \gr 2\\
 \hhline{~----}
 & \gr (12)(34) & \gr (1)(243) & \gr (142)(3) & \gr 2\\
 \hline 
(c) & \gr (12)(34) & \gr (13)(24) & \gr (14)(23) & \gr 1\\
 \hhline{~----}
 & \gr (12)(34) & \gr (14)(23) & \gr (13)(24) & \gr 1\\
 \hline 
\end{tabular}
\end{center}
\caption{Structures contributing to $\mathcal{T}_{1,2}$. The structures marked in grey give contributions
that vanish after taking derivatives.
\label{table223GUE}}
\end{table}

The case of $\T_{2,2}$ is a bit more complicated.
We are displaying the relevant permutations in Table \ref{table223GUE}.
Several choices, marked in grey, lead to results that vanish after taking derivatives. One relevant contribution arises from the factorization
$(123)=(13)(2)\cdot(12)(3)$. With two orbits inside $P$ and two orbits
inside $Q$, there are four ways to assign the cycles of $\pi$ and $\rho$ to orbits, and
one can show that the overall contribution is $-2z_{11}z_{22}-2z_{12}z_{21}$.
Further contributions arise from a different diagram, and the
two
factorizations $(12)(34)=(143)(2)\cdot(123)(4)$ and $(12)(34)=(132)(4)\cdot(134)(2)$.
Taking into account the four different ways of assigning cycles to orbits, as well as the
factor $\frac{1}{\prod_\ell v_\ell!}=\frac{1}{2}$, the contribution from each of these
choices is $z_{11}z_{22}+z_{12}z_{21}$. We thus see that all contributions sum to zero.

For $\T_{2,2}$, there is also the possibility of a partial diagonal
approximation in which one orbit from $P$ is identical to one orbit from $Q$. However,
the remaining non-identical orbits would lead to a contribution proportional to
$\T_{1,1}$, which we have already seen to be zero.

We have checked using a computer, by explicitly producing all required
factorizations, that the next   correction to $\R_4$ (with $L-V=4$) vanishes. 
There are altogether 49 diagrams with $J=1,K=3$ as well 121 diagrams with $J=K=2$,  and all contributions that do not vanish immediately after taking derivatives mutually cancel.
We refer to the appendix for     $\R_5$.

\section{Leading orders, preserved TRS}

The $2$-point spectral correlation function has been obtained semiclassically for systems
with time-reversal symmetry, but nothing of the sort has been done so far for higher
correlations. In the following, we obtain the first few orders in perturbation theory,
for the first few $\R_n$.

\subsection{$2$-point function}

As we have seen in Section II.B, the non-oscillatory part of the RMT result for $n=2$ is
\be \mathcal{R}^{\rm no}_2=1-\frac{1}{(\pi\epsilon)^{2}}+\frac{3}{2(\pi
\epsilon)^{4}}-\frac{15}{(\pi\epsilon)^{6}}+\cdots\ee
where $\epsilon=\epsilon_1-\epsilon_2$. 
Here the term proportional to $\frac{1}{\epsilon^2}$ arises from the diagonal approximation evaluated in
Section II.D, see (\ref{diag}) with a factor arising from the symmetrisation analogous to (\ref{sym}).

\begin{figure}[t] \includegraphics[scale=0.35,clip]{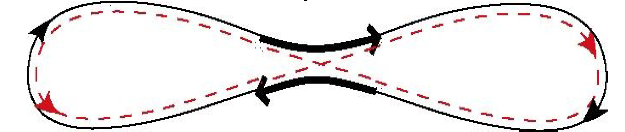}\caption{ (color
online) Sieber-Richter pair of orbits.} \label{sr}
\end{figure}

The first correction comes from the so-called Sieber-Richter pairs \cite{sieber,sieber2}, see Fig. \ref{sr}, which correspond to
the factorization $\sigma=\pi^{-1}\rho$ with  $\sigma=(12)(\bar 2\bar 1)$,
$\pi=(1\bar 2)(2\bar 1)$ and $\rho=(1\bar 1)(2\bar 2)$. The
contribution this gives to $\T_{1,1}$ is $1/(i\pi\epsilon)$, which after applying
$D_{1,1}$ leads to $i/(2\pi^3\epsilon^2)$. 
However, this result vanishes
after symmetrization (c.f. Eq.(\ref{sym})) as exchanging $\epsilon_1$ and $\epsilon_2=\eta_1$ flips the sign of $\epsilon$.
Also note that in this case symmetrization is equivalent to taking twice the real part.
However, Sieber-Richter pairs give important contributions to the spectral form factor, defined as the Fourier transform
\be
K(\tau)=\int_{-\infty}^\infty d\epsilon\,
e^{i\epsilon\tau}\widetilde \R_2(\epsilon).
\ee
This apparent
contradiction can be resolved as follows:
The asymptotic behavior of the two-point correlation function can be written as a power series \cite{longb}
\be
\R_2(\epsilon)={\rm Re}\sum_{m=2}^\infty c_m\left(\frac{1}{i\epsilon}\right)^m+\ldots
\ee
where the dots represent oscillatory terms that we are neglecting in our present approach. The $m$-th term in this series expansion is associated to the terms proportional to $\tau^{m-1}$ in the spectral form factor. However the Fourier transform has to be carried out in the complex plane, requiring to take $\epsilon$ with a small imaginary increment that can then be sent to zero. Even powers of $\tau$ are thus associated to odd powers of $\frac{1}{i\epsilon}$ that would vanish after taking the real part but give a nonzero result if the increment is included. Our present approach could   be extended to studying the asymptotics of the  spectral form factor and its equivalents for higher-order correlation functions by incorporating such imaginary increments. However we will not carry out this generalization as our emphasis is on the correlation functions themselves.

The next order, with $L-V=2$, is associated with the appropriate factorizations of
$(123)(\bar 3\bar 2\bar1)$ and of $(12)(34)(\bar 4\bar 3)(\bar 2\bar 1)$. The former has
four such factorizations, each contributing $z_{11}^2/(2\pi^2)$ to $\T_{11}$.
The latter has twenty such factorizations, each contributing $-z_{11}^2/(8\pi^2)$. Hence,
the contribution at this order to $\T_{1,1}$ is $-z_{11}^2/(2\pi^2)$. After
applying $D_{1,1}$ we get precisely $3/(2\pi^4\epsilon^4)$, in agreement with RMT.

We have checked in the computer that at the next two orders, given by diagrams with
$L-V=3$ and $L-V=4$, our semiclassical approximation also agrees with the corresponding
RMT prediction. This is checking consistency with \cite{R2b,R2c} where the spectral form factor associated to the 2-point correlation function was treated to all orders.

\subsection{$3$-point function}

As we have seen in Eq.(\ref{R3til}), the correlation function $\R_3$ may be written as a
sum containing $\widetilde\R_0=1$, $\widetilde\R_2$ and $\widetilde\R_3$. We have already
discussed how the semiclassical approximation recovers the non-oscillatory part of
$\widetilde\R_2$, so we only need to address the non-oscillatory part of
$\widetilde\R_3$. RMT prediction for this, derivable from Eq.(\ref{Pf}), is \be 
\widetilde\R^{\rm no}_3={\rm
Sym}_3\left[\frac{3}{2\pi^6}w_{12}^2w_{13}^4+\frac{1}{ \pi^6}w_{12}^3w_{13}^3+\cdots\right],\label{goe3}
\ee where we have left out terms of higher order. Here we have set
\be
w_{jk}=\frac{1}{\epsilon_j-\epsilon_k}.
\ee
We note that the indexing here is  different from $z_{jk}=\frac{1}{\epsilon_j-\eta_k}$
due to $\eta_k=\epsilon_{J+k}$, Eq. (\ref{epsilon_eta}), and it is more convenient
for writing down our final results.

Remembering that $\widetilde\R_3=2{\rm Re}\widetilde\R_{1,2}$ (see Eq. (\ref{R3real})), we focus on
$\widetilde\R_{1,2}$. Just as in the case of broken TRS, the leading order semiclassical
contribution to this quantity should come from an orbit with a single $2$-encounter,
correlated with two other orbits, see Fig. \ref{butterfly}. However, as we have already seen this does not actually
contribute anything because application of $D_{1,2}$ to a function that depends on only
two variables returns zero.

\begin{figure}[b] \includegraphics[scale=0.4,clip]{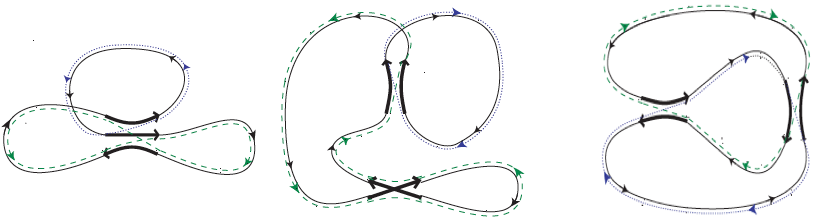}\caption{ (color
online) Diagrams contributing to $\widetilde\R_{12}$.
There is one orbit, depicted by a full line, correlated with two others depicted
by dashed lines. (Figure from the online appendix of \cite{longa}.)} \label{goe12}
\end{figure}

The second correction comes from the diagrams in Figure \ref{goe12}, having $L-V=2$. Without actually performing any
calculations, we can see that their contributions to $\T_{1,2}$ are real, but
after applying $D_{1,2}$ we arrive at an imaginary quantity. Since we need the real part
of $\widetilde\R_{1,2}$, these diagrams do not contribute either.

\begin{table} 
\begin{center}
\begin{tabular}{|c|c|c|c|c|}\hline
& $\sigma$ & $\pi$ & $\rho=\pi\sigma$ & mult.  \\ 
   
 \hline 
(a) & $(1234)$ & $(132\overline{4})$ & $(1\overline{4}\overline{1})(2)$ & 6  \\
 \hline 
(b) & $(1234)$ & $(12\overline{3}\overline{4})$ & $(1\overline{3}\overline{1})(3)$ & 3\\
 
 \hline 
(c) & $(1234)$ & $(12\overline{4}\overline{3})$ & $(1\overline{4})(24)$ & 2\\
 \hline 
(d) & $(1234)$ & $(1\overline{2}3\overline{4})$ & $(1\overline{1})(2\overline{4})$ & 2\\
 \hline  
(e) & $(123)(45)$ & $(1245\overline{3})$ & $(14\overline{3}\overline{1})(5)$ & 12\\ 
 \hline 
(f) & $(123)(45)$ & $(1342\overline{5})$ & $(1\overline{5}\overline{3}\overline{4})(3)$ & 8\\ 
 \hline 
(g) & $(123)(45)$ & $(1452\overline{3})$ & $(1\overline{3}\overline{5}\overline{1})(5)$ & 6\\ 
 \hline 
(h) & $(123)(45)$ & $(125\overline{3}4)$ & $(15)(2\overline{5}3)$ & 10\\
 
 \hline 
(i) & $(123)(45)$ & $(134\overline{2}5)$ & $(1\overline{4}24)(3)$ & 8\\
 
 \hline 
(j) & $(123)(45)$ & $(134\overline{5}\overline{2})$ & $(15\overline{5}\overline{3})(3)$ & 16\\ 
 \hline 
(k) & $(123)(45)$ & $(142\overline{3}\overline{5})$ & $(1\overline{3}\overline{4})(34)$ & 20\\ 
 \hline 
(l) & $(123)(45)$ & $(14\overline{5}3\overline{2})$ & $(1\overline{3})(34\overline{4})$ & 12\\
 \hline 
(m) & $(12)(34)(56)$ & $(16\overline{2}345)$ & $(1\overline{6}\overline{4}26)(4)$ & 192\\
 
 \hline 
(n) & $(12)(34)(56)$ & $(164\overline{2}35)$ & $(1\overline{4}26)(45)$ & 72\\
 
 \hline 
(o) & $(12)(34)(56)$ & $(1\overline{2}\overline{6}345)$ & $(1\overline{1}\overline{6}\overline{4}6)(4)$ & 96\\ 
 \hline 
(p) & $(12)(34)(56)$ & $(1\overline{2}\overline{4}356)$ & $(1\overline{1}\overline{4}45)(6)$ & 96\\ 
 \hline 
(q) & $(12)(34)(56)$ & $(14\overline{6}\overline{2}35)$ & $(16)(245\overline{4})$ & 72\\ 
 \hline 
(r) & $(12)(34)(56)$ & $(14\overline{2}\overline{6}35)$ & $(1\overline{4}6)(245)$ & 72\\ 
 \hline 
(s) & $(12)(34)(56)$ & $(1\overline{2}6\overline{4}35)$ & $(1\overline{1}6)(3\overline{6}\overline{3})$ & 96\\
 \hline 
(t) & $(12)(34)(56)$ & $(1\overline{6}\overline{2}\overline{4}35)$ & $(16)(2\overline{6}\overline{3}3)$ & 144\\ 
 \hline 
\end{tabular}
\end{center}
\caption{Structures that contribute to $\widetilde\R_{12}$ and require TRS.
See Table \ref{table124GUE} for those not requiring TRS. The table omits
structures whose contributions
 vanish after taking derivatives. It shows one representative cycle for each
pair of cycles related by time reversal.
\label{table124GOE}}
\end{table}

Finally, the diagrams responsible for the third correction have $L-V=3$.
They  include
the ones shown in Figure \ref{Fig1} and Table \ref{table124GUE},
not requiring time-reversal symmetry.  Table \ref{table124GOE} shows the remaining 20 diagrams, which do require time-reversal symmetry. We are only displaying choices of permutations whose contributions do not vanish after taking derivatives. 
Note that the cycles of $\sigma$, $\pi$ and $\rho$ come in pairs related
by time reversal; the table shows one representative cycle for each such
pair.
The contributions of these diagrams are all proportional to $z_{11}^2z_{12}+z_{11}z_{12}^2$. This has to be multiplied with the multiplicities as well as the factor
\be
\frac{2^{J+K-V}}{\prod_\ell v_\ell!}\frac{(-1)^L}{(2\pi i)^{L-V}}
\ee
arising from Eqs. (\ref{sum}) and (\ref{c_orth}), leading  to the overall
result
\be
\frac{1}{(i\pi)^3}(z_{11}^2z_{12}+z_{11}z_{12}^2).
\ee
After taking derivatives according to (\ref{defderiv}) we recover the desired
correlation function given in (\ref{goe3}). Here the factor 2 from (\ref{R3real}) and the divisor $J!K!=2$ from (\ref{sym})   mutually cancel.

\subsection{4-point function}

The non-oscillatory part of $\widetilde\R_4$ is predicted by RMT, according to Eq.(\ref{Pf}), to be
\be\label{R4}\widetilde\R_4^{\rm no}={\rm
Sym}_4\left[\frac{1}{8\pi^4}w_{12}^2w_{34}^2-\frac{3}{8\pi^6}w_{12}^4w_{34}^2\right]+\cdots,\ee where we included only terms up to order six.
Note that ${\rm Sym}_4$ sums over all permutations of indices, including permutations that leave the argument unchanged.

The term of order four, $w_{12}^2w_{34}^2/\pi^4$, comes from the diagonal approximation, 
in which the orbits are identical (or mutually time reversed) two by two.
This term involves factors $-\frac{1}{2\pi^2}w_{12}^2$ and 
$-\frac{1}{2\pi^2}w_{34}^2$ arising from the pairs according to
(\ref{diag_orth}), as well as a factor $\frac{1}{4}$ analogous to (\ref{sym})
and a factor 2 accounting for the two ways in which the $p$ and $q$ orbits can be paired.  

It turns out that $\widetilde{\R}_{1,3}$ only contributes to $\widetilde\R_4$ with 
terms of order seven, so it doesn't need to be considered in the context of the above prediction.

We still have to take into account $\widetilde{\R}_{2,2}$. This quantity allows for a partial 
diagonal approximation, see Eq.(\ref{partial}). This in fact reproduces the term of order six in (\ref{R4}). Here a factor $\frac{3}{4\pi^2}w_{12}^4$ arises from the pairs of orbits contributing to the term proportional to
$\epsilon^{-4}$ in $\widetilde\R_2$, and the factor 
$-\frac{1}{2\pi^2}w_{34}^2$ accounts for coinciding or mutually time reversed orbits. The factor $\frac{1}{4}$ from (\ref{sym}) is compensated because for the two $p$ orbits we have two choices of which is included in the diagonal pair and which is included in the pair of orbits differing in encounters, and the same choice arises for the $q$ orbits.

\begin{table} 
\begin{center}
\begin{tabular}{|c|c|c|c|c|}\hline
& $\sigma$ & $\pi$ & $\rho=\pi\sigma$ & mult.  \\ 
\hline 
(a) & $(123)$ & $(13)(2)$ & $(12)(3)$ & 1\\
   
 \hline 
(b) & $(12)(34)$ & $(134)(2)$ & $(123)(4)$ & 4\\
 \hline 
(c) & $(12)(34)$ & $(1\overline{4})(2\overline{3})$ & $(1\overline{3})(2\overline{4})$ & 1  \\
 \hline 
(d) & $(12)(34)$ & $(1\overline{2})(3\overline{4})$ & $(1\overline{1})(3\overline{3})$ & 1\\
 \hline 
\end{tabular}
\end{center}
\caption{Structures that contribute to $\widetilde\R_{22}$ for systems with TRS.
Structures whose contributions
 vanish after taking derivatives are omitted. The table shows one representative cycle for each
pair of cycles related by time reversal.
\label{table223GOE}}
\end{table}

We are left with $\widetilde{\R}^{(0)}_{2,2}$, in which no two orbits are equal. The
semiclassical approximation of order six 
is based on the diagrams in Table \ref{table223GOE} where we have included only choices of permutations that give non-vanishing contributions after taking derivatives.
However these are all proportional to $\sum_{j,j',k,k'}z_{jk}^3z_{j'k'}^3$
and thus vanish after symmetrization, as exchanging $\epsilon_j$ with $\eta_k$,
or $\epsilon_{j'}$ with $\eta_{k'}$ flips the sign.

In conclusion, the semiclassical approximation to $\R_4$ agrees with the prediction from random matrix theory, up to the sixth order in perturbation theory.

We refer to the appendix for $\R_5$.

\section{Conclusions}

We have developed a semiclassical approach that allows the calculation of the
non-oscillatory terms of arbitrary spectral correlation functions of quantum chaotic
systems. Using this approach, we have provided very strong evidence in favour of the
Bohigas-Giannoni-Schmit conjecture that all local spectral statistics of such systems are
described by those of random matrices taken from the appropriate Gaussian ensembles.

It still remains a challenge to show this agreement to all orders in perturbation theory. We believe 
it should be possible to adapt a powerful method originally introduced in
the scattering context \cite{combinat5,combinat8}, based on using and explicitly
evaluating some specific matrix
integrals that encode the semiclassical approximation. Also the connection between semiclassics and the nonlinear sigma models of RMT \cite{R2c,longb} may provide useful insight. 

Another problem still open is the semiclassical derivation of the oscillatory terms of the higher correlation
functions. This is probably amenable to treatment using the theory developed here.

SM  is grateful for support from the Leverhulme Trust
Research Fellowship RF-2013-470.
MN was supported by grants 303634/2015-4 and 400906/2016-3
from CNPq.   

\appendix

\section{5-point function } 

We want to briefly discuss the diagrams relevant for the 5-point correlation function, both for systems with and without TRS. 

\subsection{Broken TRS}

The 5-point function is determined by diagrams with $J=1$, $K=4$ as well as $J=2$, $K=3$ (which are trivially related to the diagrams where $J$ and $K$ are swapped). In either case there are no contributing diagrams for $L-V=1$ or $2$. For $L-V=3$ we obtain a large number of
diagrams which are omitted here. However their contributions 
to $\mathcal{A}_{2,3}^{(0)}$ all vanish upon differentiation and/or they include one factor $w_{jk}$ whose indices
do not appear in any of the other factors. Upon taking derivatives this factor
appears cubed. It then vanishes after symmetrization as it is odd under exchanging
indices. As $L-V=3$ corresponds to the 8th order in $w_{jk}$ this shows that all off-diagonal contributions up to this order vanish. The partial diagonal contributions vanish as well as they are based on off-diagonal contributions to $\widetilde\R_3$ which have already been shown to give zero. As desired this leaves only the diagonal approximation.

\subsection{Preserved TRS}

For time-reversal symmetric systems the RMT prediction (in leading order)
can be brought to the form\be\label{R5}\widetilde \R_5^{\rm no}={\rm
Sym}_5\left[ -\frac{3}{4\pi^8}w_{12}^2w_{13}^4w_{45}^2-\frac{1}{2\pi^8}w_{12}^3w_{13}^3w_{45}^2 \right]+\cdots.\ee 

Again all exclusively off-diagonal contributions up to 8th order in $w_{jk}$ vanish.
For $J=1$, $K=4$ the situation is exactly as without TRS, and there are no additional diagrams requiring TRS.
For $J=2$, $K=3$ there is one additional diagram with
$L-V=2$ and two structures including $\sigma=(12)(34)$,
$\pi=(1\overline 2)(34)$, $\rho=(1\overline 1)(3)(4)$. However their contributions vanish after taking derivatives. 
For $L-V=3$ there are also further diagrams, but their contributions all vanish due to either of the reasons discussed for broken TRS.

The contribution in (\ref{R5}) arises from the leading partial diagonal term.
Here three orbits are arranged according to one of the diagrams contributing to $\widetilde\R_3$, and two orbits coincide up to time reversal.
If the indices 1,2,3 are associated to the former orbits and 4, 5
to the latter orbits our previous results (\ref{diag_orth}) and (\ref{goe3})
entail factors $\frac{3}{2\pi^6}w_{12}^2w_{13}^4+\frac{1}{ \pi^6}w_{12}^3w_{13}$
and $-\frac{1}{2\pi^2}w_{45}^2$ multiplying to (\ref{R5}). There are further combinatorial factors but they cancel (a 2 to remove the $\frac{1}{2}$ from (\ref{sym}) included in the first factor, $\frac{1}{J!K!}=\frac{1}{12}$ as the factor from
(\ref{sym}) arising in the present case, 2 choices to select the $p$ orbit contributing to the diagonal approximation, and 3 choices for the $q$ orbit).

\end{document}